\documentclass[submission, Phys]{SciPost}

\usepackage{mathtools}

\hypersetup{
    colorlinks,
    linkcolor={red!50!black},
    citecolor={blue!50!black},
    urlcolor={blue!80!black}
}

\usepackage[bitstream-charter]{mathdesign}
\usepackage{enumitem}
\DeclareSymbolFont{usualmathcal}{OMS}{cmsy}{m}{n}
\DeclareSymbolFontAlphabet{\mathcal}{usualmathcal}

\usepackage[caption=false]{subfig}
\newcommand{\phantomsubfloat}[1]{
    {
        \captionsetup[subfigure]{labelformat=empty}
        \subfloat[][]{#1}
    }%
}

\begin{document}

\begin{center}{\Large \textbf{
Identifying diffusive length scales in one-dimensional Bose gases\\
}}\end{center}

\begin{center}
Frederik M{\o}ller\textsuperscript{1$\star$},
Federica Cataldini\textsuperscript{1} and
J{\"o}rg Schmiedmayer\textsuperscript{1}
\end{center}

\begin{center}
{\bf 1} Vienna Center for Quantum Science and Technology (VCQ), Atominstitut, TU Wien, Vienna, Austria

${}^\star$ {\small \sf frederik.moller@tuwien.ac.at}
\end{center}

\begin{center}
\today
\end{center}


\section*{Abstract}
{\bf
In the hydrodynamics of integrable models, diffusion is a subleading correction to ballistic propagation.
Here we quantify the diffusive contribution for one-dimensional Bose gases and find it most influential in the crossover between the main thermodynamic regimes of the gas.
Analysing the experimentally measured dynamics of a single density mode, we find diffusion to be relevant only for high wavelength excitations.
Instead, the observed relaxation is solely caused by a ballistically driven dephasing process, whose time scale is related to the phonon lifetime of the system and is thus useful to evaluate the applicability of the phonon bases typically used in quantum field simulators.
}

\vspace{10pt}
\noindent\rule{\textwidth}{1pt}
\tableofcontents\thispagestyle{fancy}
\noindent\rule{\textwidth}{1pt}
\vspace{10pt}

\section{Introduction}

The last few decades have seen significant advances in techniques for realizing and manipulating quantum many-body systems, particularly in the form of gases of ultracold atoms~\cite{RevModPhys.80.885}. 
The behavior of interacting quantum many-body systems is, however, notoriously difficult to describe~\cite{zwanzig2001nonequilibrium, Eisert2015}, spurring the development of new theoretical models.
In continuous systems with large number of particles, microscopic descriptions are generally intractable; instead, models describing the emergent, large-scale behavior of the systems are more appealing~\cite{RevModPhys.83.1405}.
For such purpose, hydrodynamics provides a powerful framework, describing the large-scale flow of densities of conserved quantities~\cite{pines1966theory, spohn2012large}.
However, integrable systems, such as the one-dimensional (1D) Bose gas, feature an infinite number of conserved quantities, thus complicating the formulation of a hydrodynamic theory; this infinite number of conservation laws poses a severe constraint of dynamics and inhibits thermalization~\cite{PhysRevLett.103.100403}.
The breakthrough came with the advent of Generalized Hydrodynamics (GHD)~\cite{castro2016emergent, bertini2016transport}, which parameterizes the hydrodynamics in terms of quasi-particles given by the Bethe Ansatz solution to integrable models. 

Hydrodynamics is an expansion of dynamics in spatial derivatives.
At lowest order in derivatives (Euler scale), currents only depend on the local densities of conserved quantities; in GHD the corresponding working equation is a collision-less Boltzmann equation~\cite{PhysRevB.97.045407}, where quasi-particles propagate ballistically at an effective velocity modified by collisions with other particles.
Beyond the Euler scale one must account for diffusive effects~\cite{PhysRevLett.121.160603}, which arise from the number of collisions experienced by the particle to be subject to equilibrium thermal fluctuations~\cite{10.21468/SciPostPhys.6.4.049}.
The result is a Gaussian broadening of the quasi-particle trajectories~\cite{PhysRevB.98.220303}.

Unlike for most non-integrable systems, diffusion in GHD arises as a subleading correction to Euler-scale hydrodynamics.
Thus, diffusive effects become evident in dynamics mainly at long time scales.
For instance, in the presence of an inhomogeneous potential, diffusion has been shown to cause a gradual thermalization of a 1D Bose gas~\cite{PhysRevLett.125.240604}.
In real systems, however, other mechanisms of integrability breaking~\cite{bastianello2021hydrodynamics}, such as noise~\cite{PhysRevB.102.161110}, losses~\cite{10.21468/SciPostPhys.9.4.044}, or violation of one-dimensionality~\cite{Mazets2008, Mazets2010, PhysRevLett.126.090602, PhysRevLett.130.030401}, are likely stronger and may therefore mask signatures of diffusion.
Indeed, in experimental tests of GHD with 1D Bose gases, no clear signs of diffusion have been observed thus far~\cite{schemmer2019generalized, malvania2020generalized}.

In this work, we study and quantify the influence of diffusion on the dynamics of one-dimensional Bose gases, particularly on time and length scales accessible in experimental setups.
First, we calculate the diffusive spreading of quasi-particle trajectories relative to their ballistic propagation velocity in thermal states.
We consider interaction strengths and temperatures ranging across several orders of magnitude, thus allowing the identification of the thermodynamic regimes where diffusive effects are most prominent.
Next, we analyse the experimental results of Ref.~\cite{PhysRevX.12.041032} in order to identify the length scales at which diffusive effects become relevant in a quasi-condensate; the setup of Ref.~\cite{PhysRevX.12.041032} is especially well-suited for such analysis, as the high degree of control over the potential enabled the excitation of a single momentum mode, thus creating excitations at a desired length scale.
Our analysis also provides a proxy for the phonon lifetime, enabling one to gauge the applicability of a phonon basis to describe dynamics.
Phonon bases have been used extensively to describe quantum field simulators realized with ultracold atoms~\cite{Gring1318, doi:10.1126/science.1257026, RauerRecurrencies, Schweigler2021, Viermann2022, Tajik2023, doi:10.1073/pnas.2301287120}; identifying the time scales at which a low-energy effective theory remains valid is thus beneficial for a number of applications.

\section{Propagation of quasi-particles in the 1D Bose gas}

Upon confinement to a single spatial dimension, an ultracold gas of $N$ repulsively interacting bosons of mass $m$ is well described by the Lieb-Lininger model~\cite{lieb1963exact, LL2}, whose Hamiltonian reads
\begin{equation}
    {\mathcal{H}} = - \sum _ { i } ^ { N } \frac { \hbar ^ { 2 } } { 2 m } \frac { \partial ^ { 2 } } { \partial z _ { i } ^ { 2 } } + g \sum _ { i < j } ^ { N } \delta \left( z _ { i } - z _ { j } \right) \; .
\end{equation}
Here, $z_i$ is the position of the $i$'th boson and $g>0$ is their coupling strength, which in an experimental system depends on the s-wave scattering length of the atoms and the potential confining the atoms to 1D~\cite{Olshanii1998}.
Often, the coupling strength is parameterized as $c = m g /\hbar^2 $, in units of inverse length.

By virtue of integrability, the 1D Bose gas can be solved exactly using the Bethe Ansatz, which identifies the elementary excitations as fermionic quasi-particles uniquely labelled by their quasi-momentum, or \textit{rapidity}, $\theta$~\cite{lieb1963exact, LL2}.
In the thermodynamics limit, the density of occupied rapidities, i.e.~the density of quasi-particles, $\rho_\mathrm{p}(\theta)$ fully characterises the thermodynamic properties of the local equilibrium macrostate.
Similarly, a density of holes $\rho_\mathrm{h} (\theta )$ can be introduced, which describes the density of unoccupied rapidities; together the two densities form the density of states $\rho_\mathrm{s} (\theta )$.
The occupational fraction of allowed rapidity states, dubbed the filling function $\vartheta (\theta) = \rho_{\mathrm{p}} (\theta)/ \rho_{\mathrm{s}} (\theta)$, equivalently characterises the macrostate.
The quasi-particle density of a thermal state at a given coupling strength and temperature is obtained by solving the Thermodynamic Bethe Ansatz (TBA)~\cite{doi:10.1063/1.1664947} (see also  Appendix~\ref{app:TBA} and Refs.~\cite{korepin_bogoliubov_izergin_1993, takahashi_1999} for more details).

Finally, the large scale dynamics of the system can be described by means of Generalized Hydrodynamics (GHD)~\cite{castro2016emergent, bertini2016transport}, which formulates a hydrodynamic theory of integrable models in terms of the propagation of its Bethe Ansatz quasi-particles.
In the following section~\ref{sec:diffusive_GHD}, we briefly review the main concepts of quasi-particle propagation and the diffusive GHD equations of the 1D Bose gas. 
For more details, see the reviews~\cite{Bouchoule_2022, DeNardis_2022}.
Next, in section~\ref{sec:quantifying}, we quantify the diffusive contribution to quasi-particle propagation in thermal states.

\subsection{Generalized Hydrodynamics at the diffuse scale} \label{sec:diffusive_GHD}

\begin{figure}
    \centering
    \includegraphics[scale=0.65]{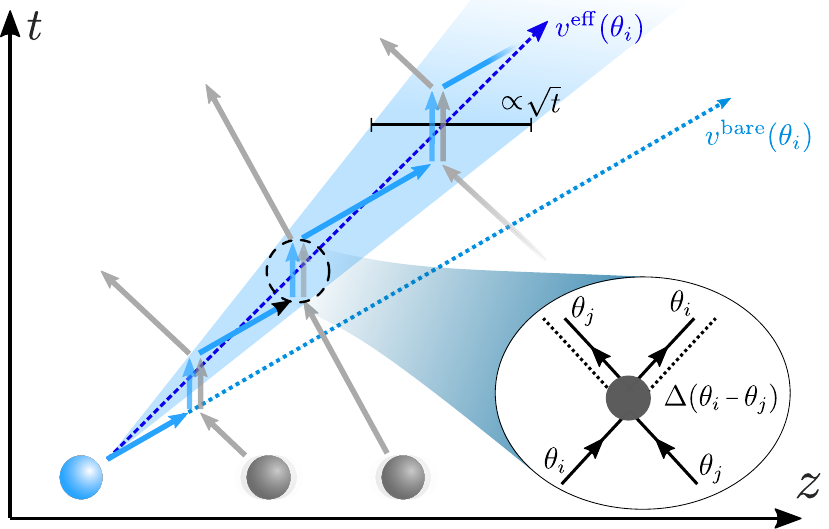}

    \caption{\label{fig:effective_velocity_diffusion}
    Illustration of quasi-particle propagation.    
    A 'tagged' quasi-particle with rapidity $\theta_i$ (blue) propagates ballistically with an average velocity $v^{\mathrm{eff}}(\theta_i)$ through a system with quasi-particle density $\rho_\mathrm{p}(\theta)$.
    The effective velocity is a modification of the bare velocity $v^{\mathrm{bare}}(\theta_i) = \hbar \theta_i /m$ following collision with other particles (grey), as each collision is associated with an apparent delay of the particles (or equivalently a positional shift $\Delta$).
    Thermal fluctuations of $\rho_\mathrm{p}(\theta)$, and thus the number of collisions experienced by the particle, result in diffusive corrections, which manifest as a  broadening of the quasi-particle trajectory on the order of $\sqrt{t}$.
    Figure inspired by Ref.~\cite{doyon2023generalized}.
    }
\end{figure}

The soliton-gas interpretation of GHD~\cite{PhysRevLett.120.045301} provides an intuitive picture of quasi-particle dynamics at different orders of the hydrodynamics expansion.

At the Euler scale, the transport of quantities is facilitated by ballistically propagating quasi-particles.
The \textit{bare} propagation velocity of quasi-particles is modified through collision with other particles, as each two-body elastic collision is associated with a Wigner time delay~\cite{PhysRev.98.145}.
Thus, as a particle with rapidity $\theta$ traverses a thermodynamic state with quasi-particle density $\rho_{\mathrm{p}}$, averaging over all the Wigner time delays experienced by the particle results in an \textit{effective} propagation velocity $v^{\mathrm{eff}}(\theta)$ (see figure~\ref{fig:effective_velocity_diffusion}).
On the quantum mechanical level, the effective velocity can be viewed as a modified group velocity of the quasi-particles, following local interactions with other particles, whereby properties, such as the momentum $p(\theta)$ and energy $\epsilon (\theta)$, of the particle $\theta$ become modified or "dressed"~\cite{PhysRevX.10.011054, Borsi_2021}.

At the next order of spatial derivatives, diffusive contributions to the dynamics are accounted for~\cite{PhysRevLett.121.160603}.
Quasi-particle diffusion arises from thermal fluctuations~\cite{10.21468/SciPostPhys.6.4.049, PhysRevB.98.220303}.
For the duration $t$, a quasi-particle will traverse a region of length $L \propto t$, through which $\rho_{\mathrm{p}}$ exhibits thermal fluctuations of order $1/\sqrt{L}$.
Thus, the number of collisions experienced by the particle, and in turn the distance it travels, fluctuates by $\sqrt{t}$~\cite{PhysRevB.98.220303}.
The result is a diffusive broadening of the quasi-particle front, as illustrated in figure~\ref{fig:effective_velocity_diffusion}.
Further, the Euler scale terms of the GHD equation are unaltered by scaling $z  \to z L$ and $t \to t L$, whilst the diffusive terms are rescaled by a factor $1/L$~\cite{PhysRevLett.125.240604}, thus highlighting diffusive effects in GHD as subleading to the ballistic quasi-particle propagation.

Assuming only large scale variations in space and time, the quasi-particle distribution of an inhomogeneous system becomes time and space dependent $\rho_\mathrm{p} = \rho_\mathrm{p} (z,t,\theta)$; the GHD equation describes the evolution of this distribution.
Note, in the following all $(z,t)$-dependencies have been omitted for a more compact notation.
On the diffusive scale, the advective form of the GHD equation (describing the evolution of the filling $\vartheta$) reads
\begin{equation}
    \partial_t \vartheta(\theta) + v^{\mathrm{eff}} (\theta) \, \partial_z \vartheta(\theta) = \frac{1}{2 \rho_{\mathrm{s}}(\theta)} \left( \mathbf{1}- \frac{\vartheta \mathbf{\Delta}}{2 \pi} \right) \: \partial_{z}\left( \left(\mathbf{1}-\frac{\vartheta \mathbf{\Delta}}{2 \pi} \right)^{-1} \rho_{s}(\theta) \: \mathbf{D} \: \partial_{z} \vartheta (\theta) \right) \; ,
    \label{eq:GHD_equation_diffusive}
\end{equation}
where the effective velocity is given by
\begin{equation}
    v^{\mathrm{eff}} (\theta) = \frac{ (\partial_\theta \epsilon)^{\mathrm{dr}} (\theta)}{ (\partial_\theta p)^{\mathrm{dr}} (\theta)} \; ,
    \label{eq:effective_velocity}
\end{equation}
with $\epsilon(\theta) = \hbar^2 \theta^2 /(2m)$ and $p(\theta) = \hbar \theta$ being the single-particle energy and momentum, respectively.
The superscript 'dr' denotes that the function has been dressed, following the equation
\begin{equation}
    g^{\mathrm{dr}} (\theta) = g(\theta) -  \frac{1}{2 \pi} \int_{-\infty}^{\infty} \mathrm{d}\theta' \; \Delta(\theta , \theta') \vartheta(\theta') g^{\mathrm{dr}}(\theta') \; ,
    \label{eq:dressing}
\end{equation}
where $\Delta (\theta, \theta') =  \frac{- 2 c}{ c^2 + (\theta-\theta')^2}$ is the two-body scattering kernel of the Lieb-Liniger model.
Further, in eq.~\eqref{eq:GHD_equation_diffusive} we have employed bold symbols to indicate integral operators and the following shorthand notation for their action 
\begin{equation}
    \left( \mathbf{K} g \right) (\theta) \coloneqq \int_{-\infty}^{\infty} \mathrm{d} \theta^{\prime} \: K\left(\theta, \theta^{\prime}\right) g\left(\theta^{\prime}\right) \; ,
\end{equation}
where $K\left(\theta, \theta^{\prime}\right)$ is the associated kernel of the generic integral operator $\mathbf{K}$.
The kernel of the integral operator $\mathbf{D}$ in eq.~\eqref{eq:GHD_equation_diffusive} reads
\begin{equation}
    D(\theta, \theta')=\frac{1}{\rho_{\mathrm{s}}^{2}(\theta)} \left( \delta(\theta-\theta') w(\theta') - W(\theta, \theta') \right) \; ,
    \label{eq:diffusion_kernel}
\end{equation}
where the off-diagonal and diagonal elements, respectively, are given by
\begin{align}
    W(\theta, \theta') &= \rho_{\mathrm{p}}(\theta)(1-\vartheta(\theta))\left[\frac{\Delta^{\mathrm{dr}}(\theta, \theta')}{2 \pi}\right]^{2}\left|v^{\mathrm{eff}}(\theta)-v^{\mathrm{eff}}(\theta')\right| \; , \\
    w(\theta) &= \int_{-\infty}^{\infty} \mathrm{d} \theta' \: W(\theta', \theta) \; .
\end{align}
The diffusive effects arise mainly from fluctuations of the state $\vartheta$, which are diagonal in rapidity and given by the expression
\begin{equation}
    \langle \delta\vartheta (\theta) \: \delta\vartheta (\theta') \rangle = \delta (\theta - \theta') \frac{\vartheta(\theta') \left( 1 - \vartheta (\theta') \right)}{\rho_{\mathrm{s}}(\theta')} \; . 
    \label{eq:filling_fluctuations}
\end{equation}
Indeed, we see that the diffusion kernel $D(\theta, \theta') $ of eq.~\eqref{eq:diffusion_kernel} is proportional to the filling fluctuations.

\subsection{Quantifying operator spreading in thermal states} \label{sec:quantifying}

According to Generalized Hydrodynamics, operator spreading in integrable systems is driven by the propagation of the Bethe Ansatz quasi-particles.
For conserved charges in particular, each quasi-particle carries a certain amount of charge density specified by the corresponding single-particle eigenvalue of the charge, which generally is a function of rapidity.
After traveling through a homogeneous system for a duration $t$, a quasi-particle with rapidity $\theta$ will ballistically have propagated a distance $\bar z(t, \theta) = t |v^{\mathrm{eff}} (\theta)| $, while the diffusive broadening of the quasi-particle trajectory is~\cite{PhysRevB.98.220303}
\begin{equation}
\begin{aligned}
    \delta \bar z^2(t, \theta) &= t \rho_\mathrm{s}^{-2} (\theta) \int_\infty^\infty \mathrm{d}\theta' \rho_\mathrm{p}(\theta') \left( 1 - \vartheta(\theta') \right) \left[\frac{\Delta^{\mathrm{dr}}(\theta, \theta')}{2 \pi}\right]^{2}\left|v^{\mathrm{eff}}(\theta)-v^{\mathrm{eff}}(\theta')\right| \\
    &=  t \rho_{\mathrm{s}}^{-2} (\theta) w(\theta) \; .
    \label{eq:quasiparticle_broadening}
\end{aligned}
\end{equation}
For the fastest quasi-particle of the system, $\bar z$ and $\delta \bar z^2$ coincide with the position and diffusive broadening of the operator spreading front, respectively.
Thus, in order to quantify the contribution of diffusion to the overall propagation of quasi-particles in the state $\rho_{\mathrm{p}}$, we introduce the measure
\begin{equation}
\begin{aligned}    
    \Gamma &= \frac{1}{n} \int_{-\infty}^{\infty} \mathrm{d}\theta \: \rho_{\mathrm{p}}(\theta) \frac{\bar z(\theta)}{ \delta \bar z^2(\theta)} \\
    &=  \frac{1}{n} \int_{-\infty}^{\infty} \mathrm{d}\theta \: \rho_{\mathrm{p}}(\theta) \rho_{\mathrm{s}}^2(\theta) \frac{ |v^{\mathrm{eff}} (\theta)|}{ w(\theta)} \; ,
    \label{eq:propagation_measure_Gamma}
\end{aligned}
\end{equation}
where $n = \int_{-\infty}^{\infty} \mathrm{d}\theta \: \rho_{\mathrm{p}}(\theta)$ is the atomic density.
Physically, $\Gamma$ describes the average distance travelled by quasi-particles relative to the variance of their trajectory, weighted according to the distribution of particles.

\begin{figure}
    \centering
    \includegraphics{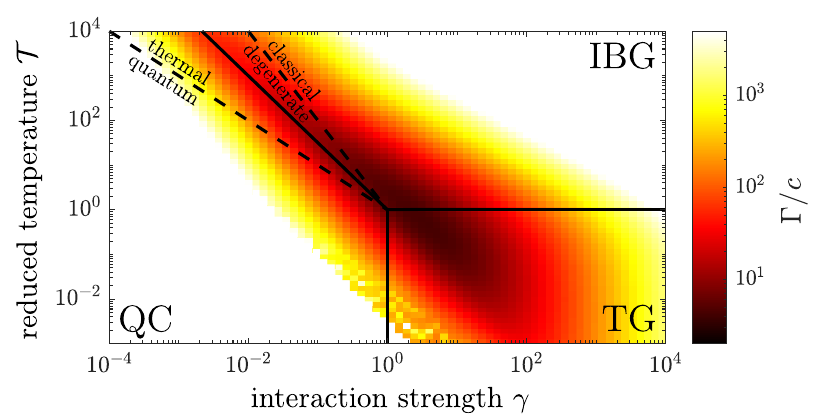}

    \caption{\label{fig:diffusion_diagram}
    Propagation metric $\Gamma$ scaled by the coupling strength $c$ for thermal states at different dimensionless temperatures $\mathcal{T}$ and interaction strengths $\gamma$.
    Lower values signify a relative increased contribution of diffusion to quasi-particle propagation.
    The solid lines indicate the boundary between the three main thermodynamic regimes of the Lieb-Liniger model, namely the quasi-condensate (QC), ideal Bose gas (IBG), and Tonks-Girardeau gas (TG).
    The dashed lines mark additional sub-regimes, namely the thermal-quantum statistics boundary for the fluctuations of the quasi-condensate and the degeneracy transition of the ideal Bose gas.
    }
\end{figure}

At thermal equilibrium, the state of a homogeneous 1D Bose gas is characterized by just two dimensionless parameters: the interaction strength $\gamma = c/n$ and the reduced temperature $\mathcal{T} = \frac{ 2 m k_\mathrm{B} T}{\hbar^2 c^2 } $, where $T$ is the temperature and $k_\mathrm{B}$ Boltzmann's constant.
These two parameters span an entire phase-diagram of the 1D Bose gas, whose regimes can be identified via the $g^{(2)}$-function~\cite{PhysRevA.71.053615}.
For thermal states with fixed values of $\gamma$, we find that $\Gamma$ scales linearly with the coupling strength $c$.
Therefore, $\gamma$ and $\mathcal{T}$ fully determine the scaled measure $\Gamma / c$ for thermal states, which we compute for a large span of parameters and plot in figure~\ref{fig:diffusion_diagram}.
For comparison, the different regimes of the Lieb-Liniger phase diagram have been indicated by solid lines.

We find that a minimum of $\Gamma/c$, indicating a maximal broadening of quasi-particle trajectories relative to their distance travelled, is located near the crossover between the three main regimes of the Lieb-Liniger model, corresponding to $\mathcal{T} \approx \gamma \approx 1$.
At higher temperatures, diffusive effects appear to be most relevant for temperatures $ \mathcal{T} \sim \gamma^{-3/2}$, which coincides with the crossover between the quasi-condensate and ideal Bose gas regimes.
Interestingly, we find that diffusion persists even at rather low temperatures; a similar observation was made in Ref.~\cite{urichuk2023navierstokes}, where diffusion at very low temperatures was shown to result in an effective viscous hydrodynamics.

Upon approaching any of the asymptotic regimes, the quasi-particle propagation becomes completely ballistic, indicated by a very large value of $ \Gamma $.
In the Tonks-Girardeau and ideal Bose gas limits this behavior is expected; free particles do not diffuse, and in these two regimes the gas behaves as free fermions and bosons, respectively.
Meanwhile, in the quasi-condensate regime, the interaction energy dominates the average energy per atom, thus making density fluctuations energetically costly, which in turn represses diffusion.
The 1D Bose gas is essentially maximally diffusive in the thermodynamic regime furthest from any of the asymptotic regimes above.
Note, that we find no direct relation between $\Gamma$ and the $g^{(2)}$-function; instead, diffusion appears maximal for kinetic and interaction energies around $E_{\mathrm{int}} \sim E_{\mathrm{kin}} \sim N k_B T/2$.
See Appendix~\ref{app:particle_energy} for more.\\

\begin{figure}
    \centering
    \includegraphics{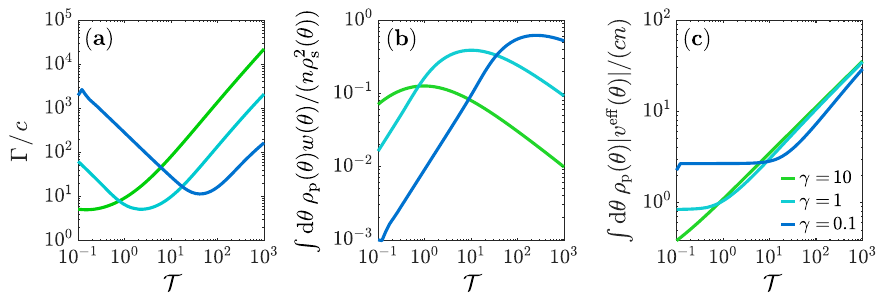}
    \phantomsubfloat{\label{fig:diffusion_diagram_cuts_a}}
    \phantomsubfloat{\label{fig:diffusion_diagram_cuts_b}}
    \phantomsubfloat{\label{fig:diffusion_diagram_cuts_c}}
    \vspace{-3\baselineskip}
    
    \caption{\label{fig:diffusion_diagram_cuts}
    Temperature scaling of diffusive and ballistic propagation.
    \textbf{(a)} Vertical cuts of the diagram~\ref{fig:diffusion_diagram} for three different interaction strengths $\gamma$.
    \textbf{(b)} Diagonal elements of the diffusion kernel weighted by the quasi-particle density: Diffusive contributions scale with increasing thermal fluctuations, until the gas starts transitioning into a regime of free particles.
    \textbf{(c)} Effective velocity weighted by the quasi-particle density. The ballistic contribution plateaus at lower temperatures as the filling function approaches a Fermi sea. 
    }
\end{figure}

To unravel the individual scaling behavior of ballistic and diffusive propagation, we plot their weighted contribution as function of temperature for different values of $\gamma$ in figure~\ref{fig:diffusion_diagram_cuts}.
For reference, figure~\ref{fig:diffusion_diagram_cuts_a} shows the corresponding values of $\Gamma / c$ (equivalent to vertical cuts of figure~\ref{fig:diffusion_diagram}). 
Starting with the diffusive contribution, plotted in figure~\ref{fig:diffusion_diagram_cuts_b}, we find that it is maximal at a temperature $\mathcal{T}$ inversely proportional to interaction $\gamma$; for lower temperatures, the diffusive contribution increase linearly with $\mathcal{T}$, while for higher temperatures it decreases again, scaling as $\mathcal{T}^{-1/2}$.
The vanishing of diffusion at very low temperatures follows from the suppression of thermal fluctuations.
Meanwhile, for very high temperatures and fixed particle number, the occupation of individual rapidity states becomes very low as quasi-particles occupy increasingly higher rapidities; such behavior is indicative of the transition towards free bosons, resulting in the eventual decrease of diffusion.

In contrast, the ballistic contribution, plotted in figure~\ref{fig:diffusion_diagram_cuts_c}, is almost constant for lower temperatures until it suddenly starts increasing proportionally to $\mathcal{T}^{1/2}$.
Approaching the ground state, the system is described by a Fermi sea of rapidity states, i.e.~the filling function assumes the value 1 between two $\gamma$-dependent Fermi point and 0 everywhere else. 
The effective velocity evaluated at the Fermi point is equal to the sound velocity of the gas~\cite{Cazalilla_2004}, while for rapidities between the two Fermi points, $v^{\mathrm{eff}}(\theta)$ is approximately a linear function.Thus, approaching lower temperatures, the fermionic statistics of the quasi-particles ensure that rapidity states up to the Fermi point remain populated, resulting in the plateau of the weighted ballistic contribution.
As the temperature increases and the Fermi sea melts, interactions in the gas become less relevant, and the effective velocity approaches $\hbar \theta / m$; for large rapidity this velocity is far greater than the elements of the diffusion kernel, thus leading to a quasi-particle propagation completely dominated by ballistic motion.\\

Finally, it should be stressed that $\Gamma$ can be computed for any state $\vartheta$, not just thermal states.
Indeed, explicitly constructing a state which maximizes fluctuations will result in a larger diffusive broadening of the quasi-particle trajectory.
An important non-thermal state, is the initial state of the seminal quantum Newton's cradle experiment~\cite{kinoshita2006quantum}; here, two halves of a (typically thermal) state have been boosted to large, opposite rapidities~\cite{Le2023}.
Notably, boosting the state substantially increases the quasi-particle velocities while merely shifting $w(\theta)$ towards higher rapidities, thus resulting in an increase in $\Gamma$.
Nevertheless, diffusive effects are important to the long time-scale dynamics of the quantum Newton's cradle, as purely ballistic propagation can produce quasi-particle distributions featuring very fine structures in the ($z,\theta$)-phase-space~\cite{10.21468/SciPostPhys.6.6.070}.
Such small wavelength features are much more susceptible to diffusive effects, which over time lead to the features being washes out and the system thermalizing~\cite{MOLLER2023112431}.

\section{Quasi-particle dynamics following a single-mode quench in a quasi-condensate}

The measure $\Gamma$ describes the relative contribution of ballistic and diffusive quasi-particle propagation in a given state.
However, the measure fails to account for the role of inhomogeneity; indeed, diffusive effects become increasingly important at shorter length scales.
In this section, we employ a linearized version of the diffusive GHD equation to analyze the dynamics of a single momentum mode of the atomic density, i.e.~an excitation with a well-defined length scale.
Such a single-mode quench was achieved in the recent experiment of Ref.~\cite{PhysRevX.12.041032}.
Although the Bose gas was realized in the quasi-condensate regime where diffusive effects are very weak, the ability to excite a single momentum mode makes such an experimental platform ideal for studying dynamics at different length scales.

\subsection{Experimental setup and protocol}

The experimental setup is already described in detail in Ref.~\cite{PhysRevX.12.041032}.
Hence, we will only summarize the details most relevant to this study.

\begin{figure}
    \centering
    \includegraphics{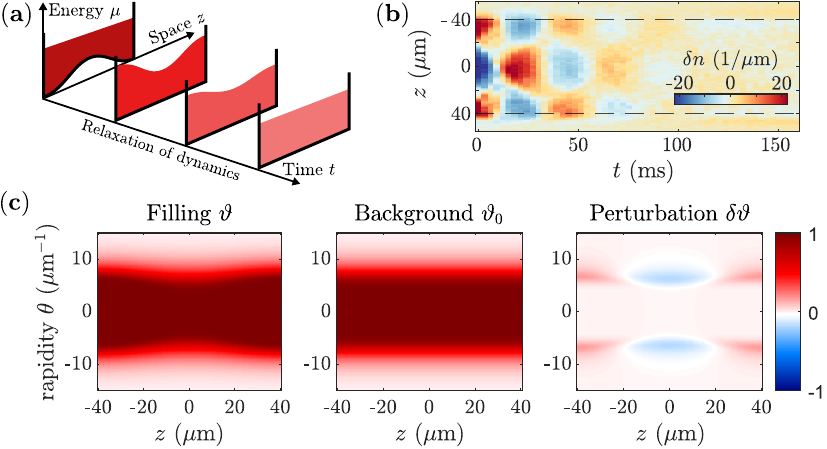}
    \phantomsubfloat{\label{fig:setup_summary_a}}
    \phantomsubfloat{\label{fig:setup_summary_b}}
    \phantomsubfloat{\label{fig:setup_summary_c}}
    \vspace{-1.5\baselineskip}
    
    \caption{\label{fig:setup_summary}
    \textbf{(a)} Illustration of the experimental protocol of Ref.~\cite{PhysRevX.12.041032}: A 1D box trap with a cosine-modulated bottom is realized using a DMD. At time $t =0$, dynamics is initiated by quenching the box bottom to flat. 
    \textbf{(b)} Measured evolution of density perturbation in quench (c). The dashed lines indicate the position of the box walls.
    \textbf{(c)} Filling function of the initial thermal state of quench (c), along with its decomposition into a stationary homogeneous background and perturbation.
    }
\end{figure}

In the experiment, a Bose gas of $^{87}$Rb atoms are trapped on an atom chip~\cite{atomchips}, which creates a strong magnetic potential along two (transverse) axes, while providing weak trapping along the third (longitudinal or 1D) axis, effectively realizing a quasi-1D gas.
A digital micromirror device (DMD) enables the creation of arbitrary optical potentials along the 1D axis~\cite{Tajik:19}; for this protocol, the atoms are initially trapped in a box potential of length $L = 80 \: \mu\mathrm{m}$. 
By modulating the bottom of the box trap in the shape of a cosine, a density perturbation in the shape of single momentum mode is imprinted in the condensate (see Fig.~\ref{fig:setup_summary_a}).
Once confined in the box, the condensate density is 60-80~$\text{atoms}/\mu\mathrm{m}$, and the initial state is well-described by a thermal state with temperature in the range 50-120~nK (depending on the degree of cooling performed).
Here we consider three quenches with dimensionless interaction strength and temperature:
\begin{enumerate}[label=(\alph*)]
\item $\gamma = 1.8 \cdot 10^{-3}$ and $\mathcal{T} = 1.2 \cdot 10^3$,
\item $\gamma = 2.1 \cdot 10^{-3}$ and $\mathcal{T} = 2.9 \cdot 10^3$,
\item $\gamma = 1.6 \cdot 10^{-3}$ and $\mathcal{T} = 2.8 \cdot 10^3$.
\end{enumerate}
Both the coupling strength and temperature of the system are acquired by independent measurement; the former is computed from the transverse trapping frequency, while the latter is measured by means of density ripples thermometry~\cite{PhysRevA.80.033604, PhysRevA.81.031610, PhysRevA.104.043305}.
Comparing with figure~\ref{fig:diffusion_diagram}, we find that the gas is realized in the quasi-condensate phase near the thermal-quantum statistics boundary for the fluctuations; in this parameter regime, diffusive effects are expected to be weak.

To initiate dynamics, the confining potential is quenched to a flat-bottomed box trap at time $t=0$ and the subsequent dynamics and relaxation is monitored by recording the atomic density via absorption imaging.
An example of the measured density evolution is plotted in figure~\ref{fig:setup_summary_b}.
Owing to a high degree of control over the 1D potential, a single density mode of the condensate can be excited, that is, the atomic density inside of the box trap is accurately given by $n (z,t) = n_0 + \delta n_j(t) \cos \left( k_j z \right)$, where $\delta n_j (t)$ is the amplitude of the mode and $k_j = 2 \pi j /L$ is its momentum.
For the quenches analysed here, only the $j = 1$ mode is excited, however, the ability to also address higher modes was demonstrated in Ref.~\cite{PhysRevX.12.041032}.
In the context of Bogoliubov theory, the excited density mode corresponds to an eigenstate of the effective low-energy Hamiltonian~\cite{PhysRevA.67.053615}.
However, as previously mentioned, the phononic basis does not represent the true eigenstates of the system~\cite{PhysRevLett.130.140401}.
Further, due to the high temperature of the condensate, the applicability of the low-energy model is very limited; we will discuss and demonstrate this in the following section.

\subsection{Analysis of the single-mode dynamics}

\subsubsection{Linearization of the diffusive GHD equation}

Simulating diffusive GHD according to eq.~\eqref{eq:GHD_equation_diffusive} generally requires powerful numerics~\cite{10.21468/SciPostPhys.8.3.041, MOLLER2023112431}.
However, following the approach of Ref.~\cite{10.21468/SciPostPhysCore.1.1.002}, one can diagonalize the GHD propagation kernel, enabling a direct study of diffusive length scales.
Here we briefly summarize the approach.

Consider the time-dependent filling function consisting of a stationary, homogeneous background $\vartheta_0$ and an evolving perturbation $\delta\vartheta$, such that
\begin{equation}
    \vartheta (z, t, \theta) = \vartheta_0 (\theta) + \delta\vartheta(z, t, \theta)  \; .
    \label{eq:filling_perturbation}
\end{equation}
Assuming a small perturbation $\delta\vartheta \ll \vartheta_0$, we can neglect interactions within the perturbation itself and only treat the interactions between the perturbation and the stationary background.
Evaluating the effective velocity and the diffusion kernel with respect to the homogeneous background filling $\vartheta_0$, means that each Fourier mode of the filling function perturbation evolves independently.
A perturbation consisting of only a single Fourier mode $\delta\vartheta(z, t, \theta) = \delta\vartheta_j(t, \theta) e^{i k_j z}$, where $k_j= 2 \pi j /L$ with $L$ being the system size, has the time-dependent solution
\begin{equation}
    \partial_t \delta \vartheta_j(t,\theta) + \int_{-\infty}^{\infty} \mathrm{d} \theta' \: \mathfrak{D}_j(\theta, \theta') \: \delta \vartheta_j(t, \theta') = 0 \; ,
    \label{eq:GHD_equation_diffusive_mode}
\end{equation}
where we have introduced $\mathfrak{D}_j$, which is kernel of the propagation operator of the $j$'th mode and is given by 
\begin{equation}
    \mathfrak{D}_j(\theta, \theta')=i k_j v_0^{\mathrm{eff}}(\theta) \delta(\theta-\theta') + \frac{k_j^2}{2} {D}_0(\theta, \theta') \; .
    \label{eq:propagation_operator}
\end{equation} 
The propagation operator is a linear integral operator with eigenstates $f_{\omega,j}(\theta)$ and corresponding eigenvalues $\lambda_{\omega, j} \in \mathbb{C}$, where the index $\omega$ enumerates the eigenvalues.
The solution to eq.~\eqref{eq:GHD_equation_diffusive_mode} can be expressed in terms of the eigenstates as
\begin{equation}
    \delta \vartheta_j(\theta, t) = \sum_\omega \eta_{\omega, j} f_{\omega, j}(\theta) e^{-\lambda_{\omega, j} t} \;,
    \label{eq:GHD_eigenmode_evolution}
\end{equation}
where the coefficients $\eta_{\omega, j}$ are determined from the initial state, such that
\begin{equation}
    \delta \vartheta(z, t=0,\theta)=\sum_j e^{i k_j z} \sum_\omega \eta_{\omega, j} f_{\omega, j}(\theta) \; .
    \label{eq:linGHD_coeffs}
\end{equation}
According to eqs.~\eqref{eq:GHD_eigenmode_evolution} and \eqref{eq:linGHD_coeffs}, each $k$-mode (Fourier mode) of the filling can be expressed as a sum of eigenstates $f_{\omega,j}(\theta)$, whose evolution is determined by their corresponding eigenvalue $\lambda_{\omega,j}$.
The imaginary part of $\lambda$ describes ballistic propagation, here manifesting as an oscillation of the eigenstate; the real part of $\lambda$ represents diffusion, which dampens the oscillation. 
Following the non-degeneracy of the spectrum of $\mathfrak{D}_j$, each eigenstate $f_{\omega}$ evolves ballistically at a different frequency and eventually dephase with respect to one-another, causing an apparent relaxation of the mode $\delta \vartheta_j$.
On shorter time scales, accounting for diffusion merely leads to (slightly) faster apparent relaxation of the mode, thus highlighting the difficulty in experimentally observing diffusion.
On long time scales, however, the redistributes of rapidities via diffusion causes the $k$-mode to relax from a decoherent superposition of waves to a stationary state.

\subsubsection{Comparing experimental observations with linearized GHD}

\begin{figure}[t]
    \centering
    \includegraphics{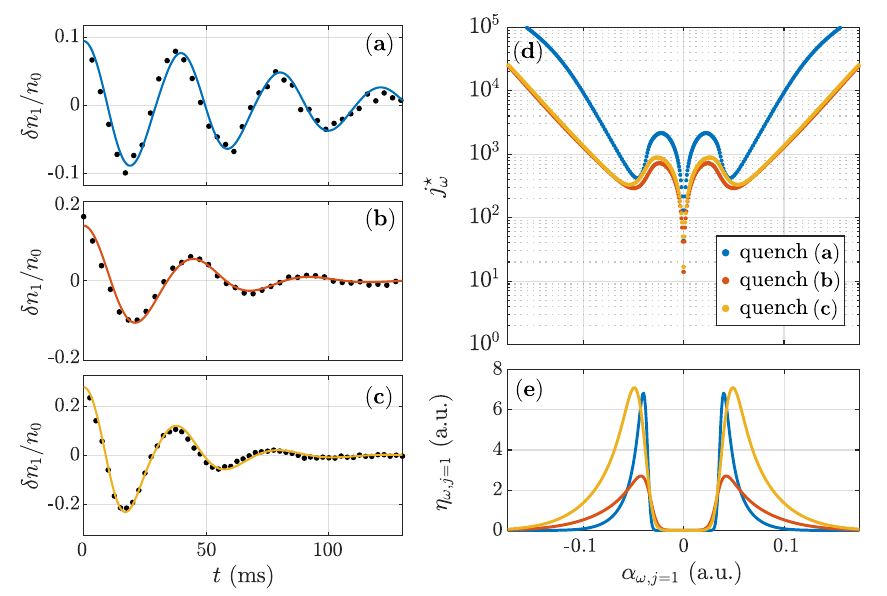}
    \phantomsubfloat{\label{fig:critical_mode_experiment_a}}
    \phantomsubfloat{\label{fig:critical_mode_experiment_b}}
    \phantomsubfloat{\label{fig:critical_mode_experiment_c}}
    \phantomsubfloat{\label{fig:critical_mode_experiment_d}}
    \phantomsubfloat{\label{fig:critical_mode_experiment_e}}
    \vspace{-3\baselineskip}

    \caption{\label{fig:critical_mode_experiment}
    Analysis of propagation in the single-mode quench experiment of Ref.~\cite{PhysRevX.12.041032}. 
    On the left, the experimentally measured evolution of the lowest cosine mode (black dots) compared with linearized GHD simulations (colored lines).
    The interaction strength and reduced temperatures of the background state of the three quenches are: 
    \textbf{(a)} $\gamma = 1.8 \cdot 10^{-3}$ and $\mathcal{T} = 1.2 \cdot 10^3$,
    \textbf{(b)} $\gamma = 2.1 \cdot 10^{-3}$ and $\mathcal{T} = 2.9 \cdot 10^3$,
    \textbf{(c)} $\gamma = 1.6 \cdot 10^{-3}$ and $\mathcal{T} = 2.8 \cdot 10^3$.
    On the right, \textbf{(d)} the critical modes $j_\omega^\star$ of eq.~\eqref{eq:critical_mode_def} for eigenstates of the propagation operator, and \textbf{(e)} the coefficients (i.e. population) of the eigenstates for the single-mode perturbation excited in the experiment.
    We label the eigenstates by the imaginary part of the corresponding eigenvalue $\alpha_{\omega, j =1}$.
    }
\end{figure}

To assess whether a linearization of the dynamics is valid for the experimental system, we simulate the dynamics of the gas using the linearized GHD~\cite{10.21468/SciPostPhysCore.1.1.002} and compare with experimental observations.
First, to extract the filling function of the background and perturbation, we spatially Fourier transform the filling of the initial thermal states and identify the $j=0$ mode as the stationary background.
See figure~\ref{fig:setup_summary_c} for an example.
The perturbation $\delta\vartheta (z, \theta)$ is comprised of all remaining modes; since density is a non-linear function of the filling, an excitation of a single-density mode generally features a perturbation $\delta\vartheta$ containing multiple modes.
Indeed, after Fourier transforming $\delta\vartheta (z, \theta)$ we find higher modes of the filling to be occupied, however, their amplitude is relative low (less than 10\% of the $j = 1$ mode), whereby we will focus solely on the evolution of $\delta\vartheta_{j=1} (t, \theta)$.

Next, given the background filling $\vartheta_0 (\theta)$ we calculate and diagonalize the propagation kernel, whereby the evolution of the perturbation can be calculated efficiently using eq.~\eqref{eq:GHD_eigenmode_evolution}.
The results are plotted in figure~\ref{fig:critical_mode_experiment} for the three single mode quenches.
We find that the linearized GHD describes the measured dynamics to high accuracy, basically reproducing the predictions of fully interacting (Euler-scale) GHD.
The agreement remains very accurate even for the higher amplitude quenches, where the density of the perturbation $\delta n_1$ is around 20\% of the background density $n_0$.
Note, the colored lines plotted in figures~\ref{fig:critical_mode_experiment_a}, \ref{fig:critical_mode_experiment_b}, and \ref{fig:critical_mode_experiment_c} show the results of the GHD simulation \textit{without} accounting for diffusion; when simulating the linearized GHD dynamics of the system \textit{with} diffusion, we find no discernible difference in the results.
Other mechanisms of relaxation through integrability-breaking processes can also be neglected: At such high temperatures, excitations in the transverse trapping potential would normally lead to thermalization of the system~\cite{10.21468/SciPostPhys.9.4.058, PhysRevLett.126.090602}, however an emergent Pauli blocking of the associated scattering events protects the one-dimensionality of the system~\cite{PhysRevX.12.041032}.
By virtue of the high chemical potential of the condensate, all low-rapidity states are completely filled, while at higher rapidities the filling features large thermal tails.
In transverse-exciting scattering events, outgoing rapidities are much smaller than incoming due to the large gain in transverse potential energy; thus, all outgoing scattering states fall within rapidity states already occupied, and the amplitude of the scattering process vanishes.

\subsubsection{Quasi-particle dephasing and applicability of phonon basis} \label{sec:phonon_basis}

With contributions of both diffusion and a dimensional crossover being negligible, the apparent relaxation of the density mode seen in figure~\ref{fig:critical_mode_experiment} must be due to the ballistic dephasing of quasi-particle trajectories.
In figure~\ref{fig:critical_mode_experiment_e} we plot the coefficients $\eta_{\omega, j = 1}$, representing the population of each propagation eigenstate for the $j=1$ perturbation.
The eigenstates are ordered by the imaginary part (ballistic contribution) of their corresponding eigenvalue $\alpha_{\omega,j=1}$.
At low values of $\alpha_{\omega,j=1}$ the population of the corresponding eigenstates is zero, as these states belongs to the background $\vartheta_0 (\theta)$.
For all quenches, the population $\eta_{\omega, j = 1}$ is peaked around the edge of the background and features long thermal tails.
We find that $\eta_{\omega, j}$ is mostly independent of the mode number $j$; instead its width/extend depends on the temperature, while its total area depends on the initial mode amplitude. 
Thus, one can immediately understand the temperature dependence of the relaxation rate: 
A larger spread in populated ballistic eigenvalues $\alpha_{\omega,j=1}$ leads to a faster dephasing of the eigenstates comprising the mode $\delta\vartheta_{j=1}$.

This picture has clear analogies to that of phonons in Bogoliubov theory for quasi condensates~\cite{PhysRevA.67.053615} (or more generally in Luttinger liquid theory~\cite{PhysRevLett.47.1840, Haldane_1981}):
For these effective low-energy Hamiltonians, collective phase-density excitations in the form of phonons are the exact eigenstates.
In analog quantum field simulators, basis of phonons have been used extensively to analyze experimental results (see e.g. Refs.~\cite{RauerRecurrencies, Schweigler2021, Viermann2022, Tajik2023, doi:10.1073/pnas.2301287120}).
However, when considering the underlying microscopic Hamiltonian (here the Lieb-Liniger Hamiltonian), phonons are superpositions of the true eigenstates and therefore do not have infinite lifetime.
In Ref.~\cite{PhysRevLett.130.140401}, the phonon lifetime was derived by expressing the phonon as a coherent superposition of Bethe particle-hole states; over time, dephasing of the particle-hole states leads to an apparent relaxation of the phonon, similarly to the filling of a single $k$-mode $\delta \vartheta_j$.

In the context of phonons, the variation of the populations $\eta_{\omega, j = 1}$ is thus an indication of the phonon lifetime.
Given the fast relaxation of the density mode (particularly in the hot realizations), a phonon basis is not particularly suitable for describing the system.
However, as the temperature decreases, the population of large $\alpha_{\omega,j=1}$ vanishes, substantially increasing the phonon lifetime.
For this particular setup, a condensate of 30~nK (half the temperature of quench~(a)) would be reasonably described by an effective low-energy Hamiltonian, even at long timescales.

\subsubsection{Calculating diffusive length scales}

The propagation operator kernel~\eqref{eq:propagation_operator} explicitly shows that ballistic propagation scales linearly with momentum $k$, whereas diffusion scales quadratically.
Thus, we can estimate the length scale at which diffusive effects become important by diagonalizing the propagation kernel $\mathfrak{D}_j$ and analysing its spectrum.
First, we write the eigenvalues of $\mathfrak{D}_j$ as
\begin{equation}
    \lambda_{\omega, j} \equiv i \alpha_{\omega, j} + \beta_{\omega, j} \; ,
\end{equation}
where $\alpha_{\omega, j}$ and $\beta_{\omega, j}$ are both real.
From the expression~\eqref{eq:propagation_operator}, we see that $\alpha_{\omega, j}$ and $\beta_{\omega, j}$ describes the ballistic propagation and diffusive propagation of the eigenstates, respectively.
In the numerical study conducted in Ref.~\cite{10.21468/SciPostPhysCore.1.1.002}, it was demonstrated that the spectrum of $\mathfrak{D}_j$ is non-degenerate and can be ordered with respect to the value of $\alpha_\omega$.
Further, the eigenvalues are, to a good approximation, smooth functions of $j$ scaling as
\begin{equation}
\begin{aligned}
    \alpha_{\omega, j} &\approx j \: \alpha_{\omega, j=1} \\
    \beta_{\omega, j} &\approx j^2 \: \beta_{\omega, j=1}
\end{aligned}
\end{equation}
Thus, in order to estimate whether a given mode $j$ propagates mainly ballistically or diffusively, we compute the ratio
\begin{equation}
    \frac{\alpha_{\omega, j}}{\beta_{\omega, j}} \approx \frac{1}{j} \frac{\alpha_{\omega, j=1}}{\beta_{\omega, j=1}} \coloneqq \frac{j_{\omega}^\star}{j} \; ,
    \label{eq:critical_mode}
\end{equation}
where we have defined the critical mode
\begin{equation}
    j_{\omega}^\star \coloneqq \frac{\alpha_{\omega, j=1}}{\beta_{\omega, j=1}} \; .
    \label{eq:critical_mode_def}
\end{equation}
Eigenstates of modes with $j > j_{\omega}^\star$ propagate mainly diffusively, while eigenstates of modes with $j < j_{\omega}^\star$ propagate mainly ballistically.

Calculating the critical mode $j_\omega^\star$ for the experimentally realized quasi-condensates, we obtain the spectra shown in figure~\ref{fig:critical_mode_experiment_d}.
We find that $j_{\omega}^\star \gg 1$ for all eigenstates, again demonstrating that the GHD dynamics of the addressed mode is entirely dominated by ballistic propagation of the quasi-particles.
From the spectrum of  $j_\omega^\star$ we can identify the length scales at which diffusion would become relevant:
First, we note that the critical mode is highly dependent on the eigenstate.
Since $\alpha_{\omega,j=1}$ denotes the ballistic contribution to quasi-particle propagation, we would expect $j_\omega^\star$ to be minimal around low $\alpha$-values.
This is indeed the case, however, the corresponding states all belong to the stationary background.
Likewise, at large $\alpha$ the critical mode is very high.
Interestingly, at intermediate values of $\alpha_{\omega,j=1}$, the spectra of $j_\omega^\star$ all feature a local minimum.
The location of the minimum is related to the states with maximal thermal thermal fluctuations (see eq.~\eqref{eq:filling_fluctuations}).
Further, comparing with the eigenstate population of the perturbation of figure~\ref{fig:critical_mode_experiment_e}, we find that the location of the minimum coincides with the most populated eigenstates; the observed dynamics of the system is thus mainly driven by the evolution of these particular eigenstates.
From figure~\ref{fig:critical_mode_experiment_d}, we find a minimum critical mode of $j_{\omega}^\star \sim 200$.
Hence, given the same experimental parameters, diffusive effects would become dominant for quenches of the $j=200$ mode and higher.
For an experimental box length of $L = 80$~{\textmu}m, the corresponding diffusive length scale is around 0.4~{\textmu}m, which is close to the healing length of the condensate.
At such length scales, higher order (e.g.~dispersive) corrections become relevant for dynamics~\cite{PhysRevB.96.220302, https://doi.org/10.48550/arxiv.2208.06614, De_Nardis_2023, moller2023whitham}.
Further, from an experimental perspective, length scales on that order are typically not resolvable by the available imaging techniques.

\subsection{Single-mode quench in the maximal diffusive regime} 

\begin{figure}[t]
    \centering
    \includegraphics{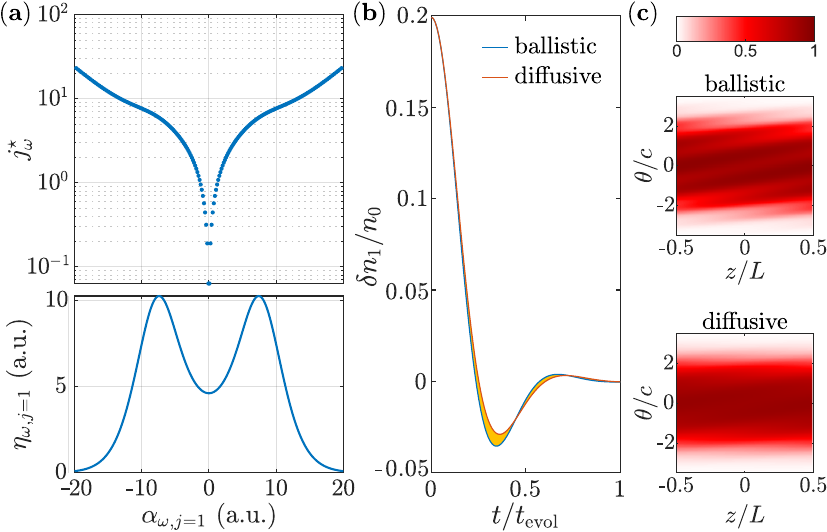}
    \phantomsubfloat{\label{fig:optimal_diffusion_setup_a}}
    \phantomsubfloat{\label{fig:optimal_diffusion_setup_b}}
    \phantomsubfloat{\label{fig:optimal_diffusion_setup_c}}
    \vspace{-1\baselineskip}

    \caption{\label{fig:optimal_diffusion_setup}
    Single-mode quench in a thermodynamic regime of maximal diffusion (see text for parameters).
    \textbf{(a)} The critical modes $j_\omega^\star$ the propagation operator and coefficients (i.e. population) of the eigenstates.
    We label the eigenstates by the imaginary part of the corresponding eigenvalue $\alpha_{\omega, j =1}$.
    \textbf{(b)} Evolution of addressed $j=1$ density mode calculated using the linearized GHD \textit{with} and \textit{without} diffusion, labelled as ballistic and diffusive, respectively. The area between the two curves has been shaded in orange for improved visibility.
    \textbf{(c)} Filling function $\vartheta$ after evolution time $t = t_\mathrm{evol}$. 
    }
\end{figure}

To illustrate the difficulty in observing diffusive effects in the 1D Bose gas from measuring its density, we consider a single-mode quench similar to the experiment~\cite{PhysRevX.12.041032} but realized near the maximally diffusive regime, identified via figure~\ref{fig:diffusion_diagram}, here $\gamma = 1$ and $\mathcal{T} = 2$.
The resulting parameters are close to a system considered in Ref.~\cite{10.21468/SciPostPhysCore.1.1.002}.
For a gas of $^{87}$Rb atoms, this would correspond to an extremely dilute system with a density of $1 \: \mathrm{atoms}/\mu\mathrm{m}$ and temperature $5.5$nK.
In order to simplify comparison to the experiment, we here also treat a quench of the $j=1$ density mode, however, we scale the length of the system $L$ to near the diffusive length scale, resulting in $L = 2\: \mu\mathrm{m}$.

The resulting critical mode spectrum and eigenstate population are plotted in figure~\ref{fig:optimal_diffusion_setup_a}.
Compared with the spectrum of the experimental system (Fig.~\ref{fig:critical_mode_experiment_d}), we find that the critical mode here is much lower for all populated states.
Further, the lack of a Fermi sea in the background state means that the highly diffusive eigenstates (low $\alpha$) are also occupied by the perturbation.
Thus, on average, we would expect modes of length scale around $L$ to exhibit diffusive behavior (following our choice of system length).

Next, in figure~\ref{fig:optimal_diffusion_setup_b}, we plot the density of the $j=1$ mode obtained from linearized GHD simulations for a duration of $t_\mathrm{evol} = 3.5 \, \mathrm{ms}$, both \textit{with} and \textit{without} diffusion.
Despite the system being realized around the maximally diffusive thermodynamic regime, the influence of diffusion on the density dynamics is very limited.
Indeed, we find a hardly measurable difference in the relaxation of the density mode upon accounting for diffusion.
Comparing higher moments~\cite{PhysRevLett.120.190601} does not yield a clear indication of diffusion either.

Finally, figure~\ref{fig:optimal_diffusion_setup_c} depicts the filling function $\vartheta$ at time $t = t_\mathrm{evol}$. 
Here, the effect of adding diffusive corrections to the ballistic propagation is evident; following ballistic propagation, the filling function develops fine structures, which are washed out in the presence of diffusion.
Upon calculating expectation values of observables, the rapidity is integrated over; functions, which are very sensitive to the exact shape of the filling, could be used to detect the presence (or lack of) of fine structure in the rapidity distribution.
However, experimental setups are highly limited in the measurable observables available.
Measurement of the rapidity density~\cite{Wilson2020} integrates over space, thus similarly hiding the underlying structure.
Hence, for a 1D Bose gas confined to a box, is it very difficult to differentiate a ballistically dephased system from one that has relaxed via diffusion.

\section{Conclusion}

To summarize, we have quantified the effect of diffusion in 1D Bose gases by calculating the diffusive spreading of quasi-particle trajectories relative to their ballistic propagation velocity.
Computing this measure for a number of thermal states, we have found that diffusive effects are most prominent in the transition regions between the different thermodynamic regimes of the Lieb-Liniger model, particularly for interaction strengths and temperatures around $\gamma \sim \mathcal{T} \sim 1$.
Diffusion, which scales with thermal fluctuations of the state, increases with temperature up to a certain point, where the gas transitions into a state of free particles.
The parameter regime of maximal diffusion is accessible by existing experimental setups, which is encouraging for future studies. 

Next, by diagonalizing the linearized propagation operator, we have identified diffusive length scales in an experimental quench of a single density mode in a quasi-condensate.
For this system, diffusive effects are found to be very weak.
Only upon approaching lengths scales around the condensate healing length, do diffusive effects become dominant.
Thus, on experimental time scales, the effect of diffusion on the density evolution can be completely neglected; the observed relaxation can instead be understood solely from the dephasing of ballistic quasi-particle trajectories.
Nevertheless, diffusive effects are still important for the dynamics at longer time scales: 
Following ballistic dephasing of the density mode, the system remains in a non-equilibrium state where ever finer structures in the filling develop~\cite{PhysRevLett.120.164101, 10.21468/SciPostPhys.6.6.070, bagchi2023unusual}; these structures are integrated over to obtain the density, and are therefore not observed in the experiment.
Once the length scale of said structures approach the critical mode, diffusive effects wash out their features and drive the system towards thermal equilibrium~\cite{PhysRevLett.125.240604, biagetti2023threestage}.
Hence, in order to observe diffusive dynamics in an experimental setup, one must realize a system where expectation values of measurable observables for balistically dephased states and thermalized states are distinguishable.
One particular setup, which fulfills this condition, is the quantum Newton's cradle.

Finally, the analysis facilitates an evaluation of the applicability of a phonon basis, which have been used extensively to describe quantum field simulators realized with ultracold gases.
For the given setup, the lifetime of the excited mode is highly dependent on temperature:
The particular quenches analyzed here feature a rather short lifetime, however, for lower temperatures (within experimental reach) the lifetime of the mode exceeds experimental time scales.

\section*{Acknowledgements}
We thank Jacopo de Nardis and Andrew Urichuk for enlightening discussions and for suggesting the measure $\Gamma$.

\paragraph{Funding information}
This work is supported by the European Research Council: ERC-AdG {\em 'Emergence in Quantum Physics'} (EmQ) under Grant Agreement No. 101097858 and the DFG/FWF CRC 1225 'ISOQUANT', (Austrian Science Fund (FWF) I~4863) and Research Unit FOR 2724 {\em'Thermal machines in the thermal world'}, (Austrian Science Fund (FWF) I~6047).-

\begin{appendix}




\section{Thermodynamic Bethe Ansatz} \label{app:TBA}

The Bethe Ansatz identifies the elementary excitations of the 1D Bose gas as fermionic quasi-particles, uniquely labelled by their quasi-momentum, or \textit{rapidity}, $\theta$~\cite{lieb1963exact, LL2}.
In the thermodynamics limit, the rapidity becomes a continuous variable, with the density of occupied rapidities, i.e.~the density of quasi-particles, $\rho_\mathrm{p}(\theta)$ fully characterising the thermodynamic properties of the local equilibrium macrostate.
Similarly, a density of holes $\rho_\mathrm{h} (\theta )$ can be introduced, which describes the density of unoccupied rapidities; together the two densities form the density of states $\rho_\mathrm{s} (\theta )$, which obeys the relation
\begin{equation}
    \rho_\mathrm{s} (\theta ) = \rho_\mathrm{p} (\theta )+\rho_\mathrm{h} (\theta ) = \frac{1}{2 \pi} -  \frac{1}{2 \pi} \int _{-\infty }^\infty d\theta ^\prime \, \Delta (\theta, \theta') \rho_\mathrm{p} (\theta ^\prime )  \; ,
\end{equation}
where $\Delta (\theta, \theta') = - \frac{2 c}{ c^2 + (\theta-\theta')^2}$ is the two-body scattering kernel of the Lieb-Liniger model.
Given $\rho_\mathrm{p}$ one can compute coarse-grained expectation values of local observables, in particular the atomic density following $n = \int_{-\infty}^{\infty}\mathrm{d}\theta \: \rho_\mathrm{p}(\theta)$. 
Equivalently, one can encode the thermodynamic properties of the system in the filling function $\vartheta (\theta) = \rho_{\mathrm{p}} (\theta)/ \rho_{\mathrm{s}} (\theta)$, which describes the occupational fraction of allowed rapidity states.
Quasi-particles of the Lieb-Liniger model follow Fermionic statistics, whereby the filling assumes values between 0 and 1.
The ground (zero temperature) state is given by a Fermi sea of rapidities, where $\vartheta(\theta) = 1$ for rapidities within the interval of two Fermi points (whose value is determined by $\gamma$) and 0 everywhere else.
Meanwhile, the filling of a thermal state is given by 
\begin{equation}
    \vartheta (\theta) = \frac{1}{1 + e^{ \varepsilon(\theta) \beta }} \; ,
\end{equation}
where $\beta$ is the inverse temperature, and the pseudo-energy $\varepsilon(\theta)$ is acquired from solving the Yang-Yang equation~\cite{doi:10.1063/1.1664947}
\begin{equation}
    \varepsilon(\theta) = \frac{\hbar^2 \theta^2}{2 m} - \mu + \frac{1}{2 \pi \beta} \int_{-\infty}^{\infty} \mathrm{d}\theta' \: \Delta(\theta , \theta') \ln \left( 1 + e^{ \varepsilon(\theta') \beta } \right) \; ,
\end{equation}
where $\mu$ is the local chemical potential.

\section{Quasi-particle propagation versus energy} \label{app:particle_energy}
\begin{figure}
    \centering
    \includegraphics{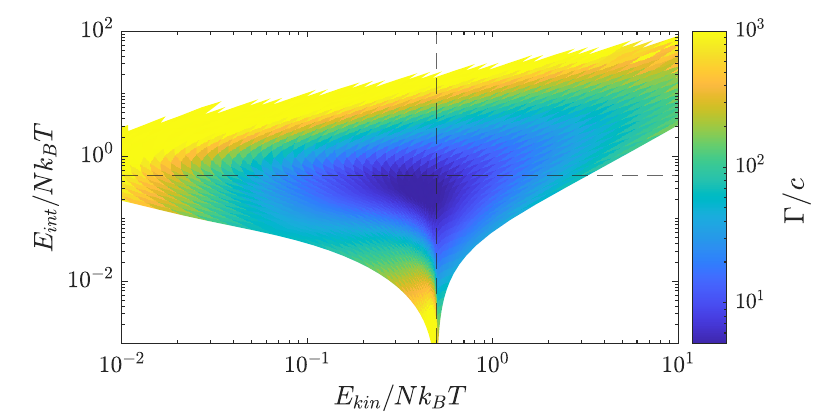}

    \caption{\label{fig:diffusion_energy}
    Propagation measure $\Gamma$ of eq.~\eqref{eq:propagation_measure_Gamma} scaled by the coupling strength $c$ as function of the mean kinetic and interaction energy per atom in the gas.
    The dashed lines mark $E_{kin}/(N k_B T) = 0.5$ and $E_{int}/(N k_B T) = 0.5$.
    }
\end{figure}

The thermodynamic regimes of the Lieb-Liniger model are identified by their two-point correlation function $g^{(2)}(0)$~\cite{PhysRevA.71.053615}.
However, although the minimum of the propagation measure $\Gamma$ (corresponding to a relative maximal influence of diffusion on quasi-particle propagation) aligns with the crossover between the different regimes, we find no clear relation between it and the $g^{(2)}$-function.
Instead, we study the relation between $\Gamma$ and the distribution of energy in the gas, which itself it related to the $g^{(2)}$-function.
In the asymptotic regimes of the Lieb-Liniger model, the system energy is dominated either by kinetic energy (free particle regimes) or interaction energy (quasi-condensate regime).
To study the relation between energy and diffusion, we plot $\Gamma/c$ as function of the mean interaction and kinetic energy per atom of the system, which, respectively, are given by~\cite{PhysRevA.85.031604}
\begin{align}
    E_{int}/N &= \frac{\hbar^2}{2m} c n g^{(2)}(0) \; ,\\
    E_{kin}/N &= E/N - E_{int}/N \; ,
\end{align}
where the total energy per atom $E/N$ is
\begin{equation}
    E/N = \frac{1}{n} \int_{-\infty}^{\infty} \mathrm{d}\theta \: \epsilon^{\mathrm{dr}} \vartheta (\theta) \; . 
\end{equation}
The result can be seen in figure~\ref{fig:diffusion_energy}, with the minimum of $\Gamma/c$ being situated around the equipartition value $E_{int}/(N k_B T) \sim E_{kin}/(N k_B T) \sim 0.5$.

\end{appendix}



\bibliography{references.bib}

\nolinenumbers

\end{document}